\documentclass[submission, Phys]{SciPost}
\usepackage{graphicx}
\usepackage{amsbsy,gensymb}
\usepackage{amsmath}
\usepackage{bm}
\usepackage{amsfonts}
\usepackage{cancel}
\usepackage{multirow}
\usepackage{amssymb}
\usepackage{xcolor}
\usepackage{float}
\usepackage{colortbl}
\usepackage{hyperref}
\usepackage{subfig}
\usepackage[export]{adjustbox}
\usepackage{appendix}
\usepackage{array}
\newcolumntype{L}{>{\arraybackslash}m{0.25\columnwidth}}
\newcolumntype{T}{>{\arraybackslash}m{0.45\columnwidth}}

\begin{document}

\begin{center}{\Large \textbf{
Supervised learning of few dirty bosons with variable particle number
}}\end{center}

\begin{center}
P. Mujal\textsuperscript{1,2*},
A. Martínez Miguel\textsuperscript{1,2},
A. Polls\textsuperscript{1,2},
B. Juli\'{a}-D\'{i}az\textsuperscript{1,2} and
S. Pilati\textsuperscript{3}
\end{center}

\begin{center}
{\bf 1} Departament de F\'{i}sica Qu\`{a}ntica i Astrof\'{i}sica,
Universitat de Barcelona, Mart\'{i} i Franqu\`{e}s 1, 08028 Barcelona, Spain
\\
{\bf 2} Institut de Ci\`{e}ncies del Cosmos (ICCUB), Universitat 
de Barcelona, Mart\'{i} i Franqu\`{e}s 1, 08028 Barcelona, Spain
\\
{\bf 3} School of Science and Technology, Physics Division, 
Universit{\`a}  di Camerino, 62032 Camerino (MC), Italy
\\
* peremujal@gmail.com
\end{center}

\begin{center}
\today
\end{center}


\section*{Abstract}
{\bf
We investigate the supervised machine learning of few interacting bosons in optical speckle disorder
 via artificial neural networks. The learning curve shows an approximately universal
power-law scaling for different particle numbers and for different
interaction strengths. We introduce a network architecture that can be trained and tested on heterogeneous datasets
 including different particle numbers.
This network provides accurate predictions for all system sizes included in the training set
and, by design, is suitable to attempt extrapolations to (computationally challenging) larger sizes.
Notably, a novel transfer-learning strategy is implemented, whereby the learning of the larger systems is substantially accelerated and made consistently accurate by including in the training set many small-size instances. 
}

\vspace{10pt}
\noindent\rule{\textwidth}{1pt}
\tableofcontents\thispagestyle{fancy}
\noindent\rule{\textwidth}{1pt}
\vspace{10pt}

\section{Introduction}
%
Supervised machine-learning is emerging as a potentially disruptive technique 
to accurately predict the properties of complex quantum systems. It has already 
allowed researchers to drastically speed-up various important computational 
tasks in quantum chemistry and in condensed-matter 
physics~\cite{carleo2019machine,carrasquilla2020machine}, including: molecular 
dynamics simulations~\cite{blank1995neural,behler2007generalized,behler2011neural,bartok2017machine,zhang2018end}, electronic structure calculations~\cite{snyder2012finding,li2016understanding,brockherde2017bypassing,ryczko2019deep,PhysRevLett.125.076402,custodio2019artificial}, structure-based molecular design~\cite{hansen2013assessment,schutt2014represent,hansen2015machine}, and protein-molecule binding-affinity predictions~\cite{ballester2010machine,khamis2015machine,jimenez2018k}.
The deep neural networks represent the most powerful and versatile  
models. In principle, they can approximate any continuous function with 
arbitrary accuracy~\cite{funahashi1989approximate}. However, training them without 
over-fitting requires extremely copious datasets, often comprising hundreds of 
thousands of training instances. Generating such datasets for large quantum systems 
is computationally impractical, unless one accepts (sometimes unreliable) approximations 
such as, e.g., density functional theory. This represents a critical problem that 
hampers the further development of machine-learning techniques for quantum systems.

A possible approach to circumvent the above problem is to adopt a transfer-learning 
strategy, as often done in the field of image analysis~\cite{caruana1997multitask}. 
In the case of quantum systems, transfer learning can be implemented by scaling to 
larger sizes the neural networks that have been trained on smaller -- therefore, 
computationally tractable -- systems. In fact, a form of size scalability is 
currently being employed in the field of molecular dynamics simulations; in 
that approach, the ground-state energies are computed as the sum 
of 
single-atom contributions, but taking into account only the short-range atomic 
environments (see, e.g., Ref.~\cite{behler2016perspective}). Proper scalability has 
recently been implemented in a few distinct ways: i) assuming the extensitivity 
property, using properly constructed size-extensive networks~\cite{mills2019extensive}; 
ii) adopting normalized descriptor-vectors of fixed size (i.e., independent of the 
physical system size)~\cite{jungsize}; iii) implementing scalable convolutional 
networks via global pooling layers, for systems of variable spatial 
extent~\cite{saraceni2020scalable}. To the best of our knowledge, statistical 
models that accept the particle number as an explicit system descriptor have not 
been investigated yet.

In this article, we consider the supervised learning of interacting bosons in a 
one-dimensional random external field. Our main goals are to quantify the 
learning speed~\cite{faber2017prediction}, in terms of prediction accuracy versus 
number of instances in the training set, and to implement flexible neural networks 
that can address different particle numbers simultaneously. The Hamiltonian we 
focus on is realistic and describes experiments performed with ultracold atoms 
in optical speckle fields~\cite{aspect2009anderson}. It represents a challenging 
computational task, belonging to the family of dirty boson problems~\cite{fisher1989boson,PhysRevB.80.104515,PhysRevLett.98.170403}. 
Recently, this model has been addressed in a study on the stability of the 
Anderson localization phenomenon against inter-particle 
interaction~\cite{PhysRevA.100.013603}. The model was shown to host a many-body 
localized phase. Here, we analyse how many instances are needed to train deep 
neural networks to accurately predict its ground-state energy, depending on the 
interaction strength and on the particle number. The training and the test sets 
are produced via an exact diagonalization technique. This choice allows us to 
avoid the common approximations employed in most supervised-machine learning 
studies. However, it limits our analysis to small particle numbers, specifically, 
up to four bosons.

\begin{figure}[t]
\centering
\includegraphics[width=0.7\columnwidth]{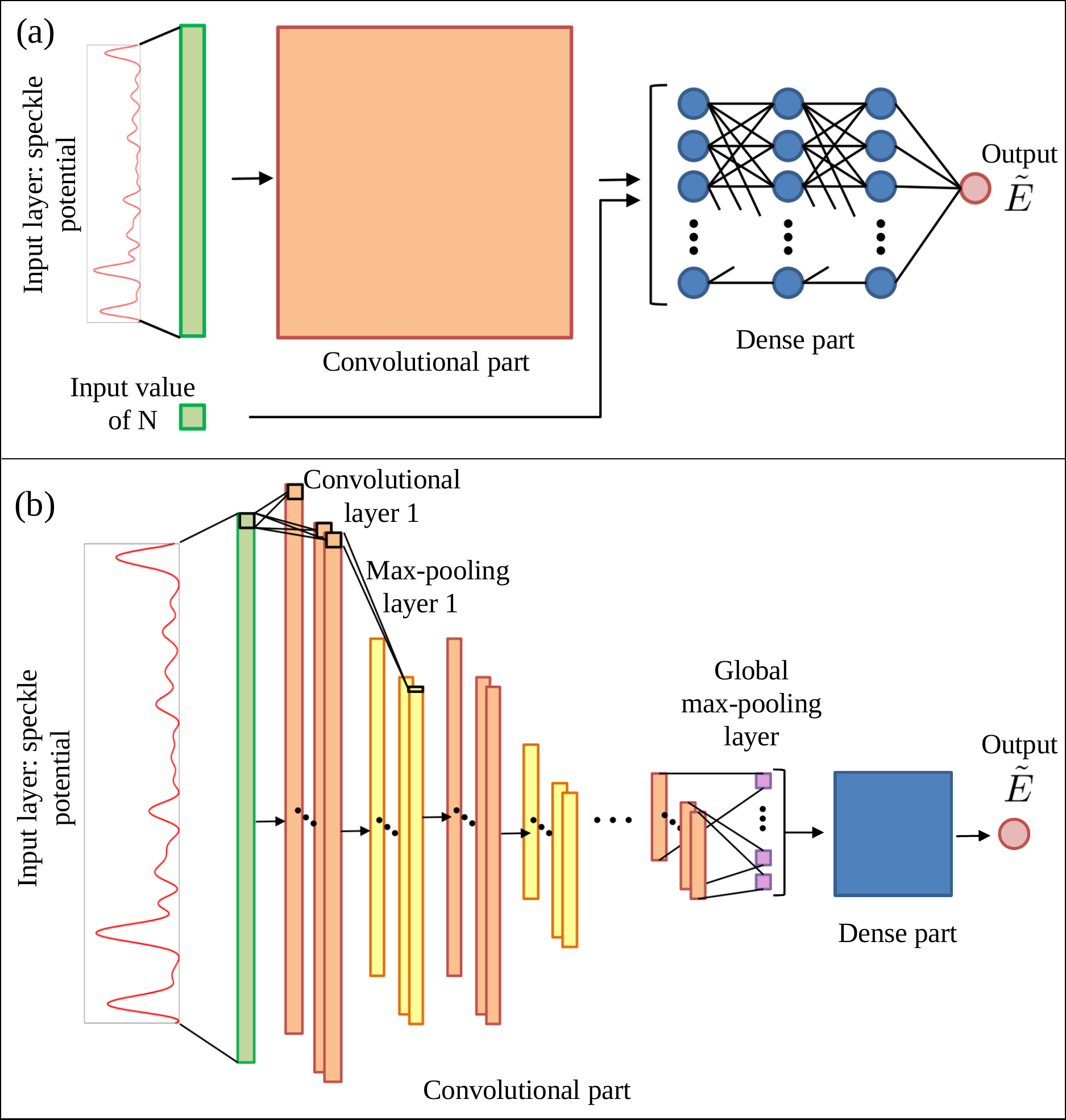}
\caption{(a) Schematic representation of the deep feed-forward neural 
network used to predict the ground-state energy, $\tilde{E}$, of few-boson 
systems (output). The input descriptors are the values of the speckle 
potential on a fine discrete grid. In the case of training with 
heterogeneous datasets, an additional system descriptor is included, 
representing the particle number $N$. This descriptor is connected 
directly to the dense part of the network. (b) Structure of the 
convolutional part of the neural network. This model is used when training on 
homogeneous datasets including instances with a unique particle number.}
\label{fig1}
\end{figure}

Notably, we implement a neural network for continuous-space quantum systems 
with variable particle number. This network combines the scalable convolutional 
architecture of Ref.~\cite{saraceni2020scalable} [see Fig.~\ref{fig1}, panel (b)], 
which can address disordered systems of variable spatial extent (but fixed particle 
number), with an additional descriptor representing the particle number. This 
descriptor bypasses the convolutional layers and is fed directly to the final 
dense layers [see Fig.~\ref{fig1}, panel (a)]. As we demonstrate here, this network 
is able to accurately predict the ground-state energies of systems with different 
particle numbers, even when considering heterogeneous datasets including instances 
with different size. The learning speed appears to be independent of the 
particle number and of the interaction strength. In fact, the prediction accuracy 
follows an approximately universal power-law scaling with the number of instances 
in the training set. Our neural network can also be used to attempt extrapolations to particle numbers larger than those included in the 
training set. The extrapolation accuracy depends on the interaction strength, and it improves if the training sets is copious and includes (relatively) large particle numbers. Furthermore, we show that the learning of the larger sizes can be 
substantially accelerated if a  training set with many small-size instances, 
which are computationally accessible, is merged with a small amount of 
instances for the larger particle number. This strategy provides consistently accurate predictions, 
also for the larger size, with a computationally feasible training set. It 
represents an alternative transfer-learning technique, paving the way to a 
novel approach to accurately predict the properties of complex quantum 
systems, for which copious training sets cannot be generated in feasible 
computational times.

Our study begins with an analysis of the learning speed, considering both 
datasets with a unique particle number, as well as the combined learning with 
heterogeneous datasets. Then, we analyse the accuracy of the extrapolations 
to particle numbers larger than those included in the training set, as well 
as the accelerated learning of relatively large systems using data for 
smaller sizes. In detail, the rest of the article is organized as follows:
the physical system we address and the computational method we employ to 
determine its ground-state energy are provided in Section~\ref{sec2},  together with a 
description of the artificial neural network introduced in this article 
and some details on the training algorithm.
The analysis on the learning speed of the few-boson problem is reported 
in Section~\ref{sec4}. Section~\ref{sec5} reports the analysis on the 
extrapolation procedure and on the accelerated learning. The summary of our 
main findings and some future perspectives are reported in Section~\ref{sec6}.
%
%
\section{Model and Methods} 
\label{sec2}

\subsection{Physical system: few 1D dirty bosons}

We consider a one-dimensional system of few repulsively interacting bosons 
in the presence of an external disordered potential. This model has been 
experimentally engineered and describes ultracold atoms subjected to 
optical speckle fields and confined in cigar-shaped traps. Specifically, 
it corresponds to the setup of early cold-atom experiments on the 
Anderson localization phenomenon~\cite{roati2008anderson,billy2008direct}. 
The Hamiltonian of the system reads
\begin{equation}
\label{Eqhamiltonian}
\mathcal{H}=\sum_{i=1}^N \left(-\frac{\hbar^2}{2m}\frac{\partial^2}{\partial x_i^2}+V(x_i)\right)
+\sum_{i<j}^N v(x_i,x_j),
\end{equation}
where $m$ is the particle mass, $N$ the number of particles, and $x_i$ 
corresponds to the position of particle $i$, with $i=1,\dots,N$. The 
two-body interaction is described by a contact interaction potential, 
\begin{equation}
v(x_i,x_j)=g\delta(|x_i-x_j|)\,,
\end{equation}
where $g$ is the parameter that defines the interaction strength. Its 
sign determines the character of the interaction: repulsive for $g>0$ 
and attractive for $g<0$. We focus on the repulsive case.

The external potential $V(x)$ represents the effect of optical speckle 
fields on ultracold atoms. It can be generated on a discrete spatial 
grid with fine spacing via the stochastic numerical algorithm described in detail 
in Refs.~\cite{ley1989specklhunte,PhysRevA.73.013606}. We produce many instances 
of speckle potentials using different pseudo-random numbers. All instances are 
characterized by the same spatial correlation length, indicated in the following 
as $\ell$, and by the same average intensity $V_0$. The correlation length allows 
one to define a characteristic energy scale, namely, the correlation energy 
$E_c=\hbar^2/(m\ell^2)$. In the following, we consider speckle fields of fixed 
spatial extent, namely, $L=20\ell$, with hard-wall boundary conditions.
The spatial grid for the speckle potential includes $1024$ points, 
corresponding to a grid spacing $\delta x \simeq 0.153\ell$. With 
such a fine grid, discretization effects are negligible.
The disorder strength is fixed at $V_0=5E_c$. Different values of the interaction 
parameter $g$ are considered; they are expressed in the following in units of 
$\hbar^2/(\ell m)$. Specifically, from now on we consider the weak interaction $g=0.05$, an intermediate value $g=0.26$, and the strong-coupling case $g=1$.

We train deep neural networks to predict the ground-state energies of 
different instances of the Hamiltonian~\eqref{Eqhamiltonian}. These 
energies are computed by means of the exact-diagonalization method 
described in Ref.~\cite{PhysRevA.100.013603}. This method is based 
on a second-quantization formalism. The Fock space of the $N$ bosons is 
built using the basis of the single-particle eigenstates of the kinetic 
energy operator. The diagonalization is performed in a truncated space 
including only the Fock basis states with kinetic energy smaller than a 
chosen threshold, following the technique introduced 
in Ref.~\cite{plodzien2018numerically}. This energy threshold determines 
both the dimension of the truncated $N$-boson Fock space $D_{\rm MB}$, and 
the required number of single-particle basis states $M$. Further details 
on the computational technique we employ are reported in 
Ref.~\cite{pmtphdthesis}. The energy thresholds we adopt in this article 
lead to the following truncation parameters: for $N=1$, we have 
$M=D_{\rm MB}=100$; for $N=2$, $M=100$ and $D_{\rm MB}=3914$; 
for $N=3$, $M=100$ and $D_{\rm MB}=88106$; and for $N=4$, $M=80$ and 
$D_{\rm MB}=552099$. 
The computational resources available to us allow producing datasets 
including different numbers of instances; specifically, we 
produce $600000$, $50000$, $2000$, and $270$ 
instances for $N=1,2,3, \mathrm{and} \ 4$, respectively, for each of the 
three values of the interaction parameter $g$ we consider. These 
datasets are available at~\cite{zenodo}.


%
\begin{table}[t]
\centering
\begin{tabular}{|L|L|T|}
\hline
\cellcolor[gray]{0.96}Layer name   & \cellcolor[gray]{0.96}Layer Function & \cellcolor[gray]{0.96}Layer description \\ \cline{1-3}
Input layer 1& Input & The value of the speckle potential in 1024 points     \\ 
\hline
Convolutional layers& ReLu & 50 filters, kernel\_size=5, strides=1, padding=same, activation=relu  \\
\hline
Local max-pooling layers& Max-pooling & Local pooling with pool\_size=3  \\ 
\hline
Global max-pooling layer& Global max-pooling  & 50 neurons = number of filters in the preceding convolutional layer\\ \hline
Input layer 2& Input &  $N$, number of particles\\\hline
Dense layers& ReLu & 30 neurons, activation=relu   \\ \hline
Output layer& Identity & 1 neuron, activation=identity   \\ \cline{1-3}
\end{tabular}
\caption{Details of the layers constituting the convolutional and the dense parts of our neural network. Definitions are standard, see for instance Ref.~\cite{nielsen2015neural}. }
\label{tablenetwork}
\end{table}
%

\subsection{Network architecture}

In Ref.~\cite{saraceni2020scalable}, deep feed-forward neural networks have 
been employed in the supervised learning of the ground-state energy of the 
Hamiltonian~\eqref{Eqhamiltonian}. However, that study addressed only the 
single particle case, namely, the case $N=1$. A scalable architecture was 
implemented using standard convolutional layers connected to dense hidden 
layers (i.e., with all-to-all connectivity) via a global pooling operation. 
This allows the model to address disordered systems of arbitrary spatial 
extent $L$. Our goal is to further develop that architecture so that it 
can address also an arbitrary particle number $N$.

Our investigation first addresses homogeneous datasets including 
instances with a single particle number, either $N=1$, $N=2$, or $N=3$. 
For this purpose, the architecture of Ref.~\cite{saraceni2020scalable} 
(represented in panel (b) of Fig.~\ref{fig1}) can be employed without 
modifications.
The system instances are represented by 1024 descriptors corresponding 
to the speckle potential intensities $V(x_k)$ on the spatial grid 
$x_k=k\delta x$, with $k=0,...,1023$. 
Since the grid spacing $\delta x$ is much smaller than the disorder 
correlation length $\ell$, these 1024 descriptors provide an 
exhaustive representation of the speckle potential of each instance.
The 1024 descriptors are fed to the convolutional part of the architecture. 
This part includes six convolutional layers with 50 filters, each followed by 
a local pooling layer. The output of the convolutional part is forwarded to 
the first of three dense layers, each including 30 neurons, via a global 
pooling layer. The final layer includes a single neuron. Its activation 
should correspond to the ground-state energy.
Thanks to the global pooling layer, this architecture can be applied to 
systems with different spatial extent $L$ (and, hence, different 
numbers of descriptors), without re-training.
To address heterogeneous datasets containing instances with different particle 
numbers, we have to extend the architecture shown in panel (b) of Fig.~\ref{fig1}. 
Specifically, we include an additional descriptor whose value corresponds to 
the particle number $N$. The corresponding neuron is linked directly to 
the first dense layer, bypassing the convolutional and the pooling layers 
(see panel (a) of Fig.~\ref{fig1}).
In principle, this should allow the model to learn how the ground-state energy 
depends on the particle number, providing predictions for arbitrary $N$. In 
Sections~\ref{sec4} and~\ref{sec5} we quantify if and to what extent this goal is achieved.
All details of the neural-network structure are reported in Table~\ref{tablenetwork}. 

\subsection{Training procedure}

The training is performed by minimizing the mean squared error (MSE), 
defined as:
\begin{equation}
    \mathrm{MSE} = \frac{1}{N_\mathrm{train}}
    \sum_{t=1}^{N_\mathrm{train}} \left(\tilde{E}_{t}-E_t \right)^2\,,
\end{equation}
where $E_t$ and $\tilde{E}_t$ are, respectively, the exact and the predicted 
ground-state energies of the training instance $t$. $N_\mathrm{train}$ 
is the number of instances included in the training set.
The optimization of the neural-network weights and biases is performed 
using the \textit{Adam} algorithm~\cite{kingma2014adam}, as implemented in 
the \textit{Keras} python library~\cite{chollet2015keras}. 
An early stopping criterion is adopted. It is based on the MSE of a validation 
set (distinct from the training set). The optimal network parameters obtained 
throughout the training process are retained. 
To quantify the accuracy of the predictions provided by the trained networks 
we consider two figures of merit. The first is the mean absolute error (MAE), 
defined as:
\begin{equation}
    \mathrm{MAE} = \frac{1}{N_\mathrm{test}}\sum_{t=1}^{N_\mathrm{test}} \left|\tilde{E}_{t}-E_t \right|\,.
\end{equation}
The second is the coefficient of determination, defined as:
\begin{equation}
    \mathrm{R^2} = 1-\frac{\sum_{t=1}^{N_\mathrm{test}} \left(\tilde{E}_{t}-E_t \right)^2}{\sum_{t=1}^{N_\mathrm{test}} \left(E_t - \left<E\right> \right)^2}\,.
\end{equation}
Here, $N_\mathrm{test}$ is the number of instances in the test set, and 
$\left\langle E\right\rangle$ is their average ground-state energy.
We stress that the instances included in the test set are distinct from 
those used for training and for validation.
It is worth recalling that perfect predictions correspond to the score $R^2=1$, while a constant function predicting the correct average $\left\langle E\right\rangle$ corresponds to the score $R^2=0$.
For the results reported in the following sections, unless otherwise specified, 
$20\%$ of the datasets are used for testing. The 
remaining $80\%$ is divided into the training data, accounting a $75\%$, and 
validation data, corresponding to the remaining $25\%$. 
It is worth mentioning that 
the data reported in this article are obtained without regularization techniques. 
In our analysis, we inspect for the occurrence of overfitting by monitoring the discrepancy between the MAE of the training set and of the validation set. In general, we find similar results, apart for the smallest training sets, for which the validation MAE is, in the worst case, up to twice the training MAE. Anyway, tuning the regularization parameter with a standard L2 regularization~\cite{nielsen2015neural} does not significantly improve the accuracy on the test set.

\section{Learning the few-body problem}
\label{sec4}

\begin{figure}[t!]
\centering
\includegraphics[width=0.65\columnwidth]{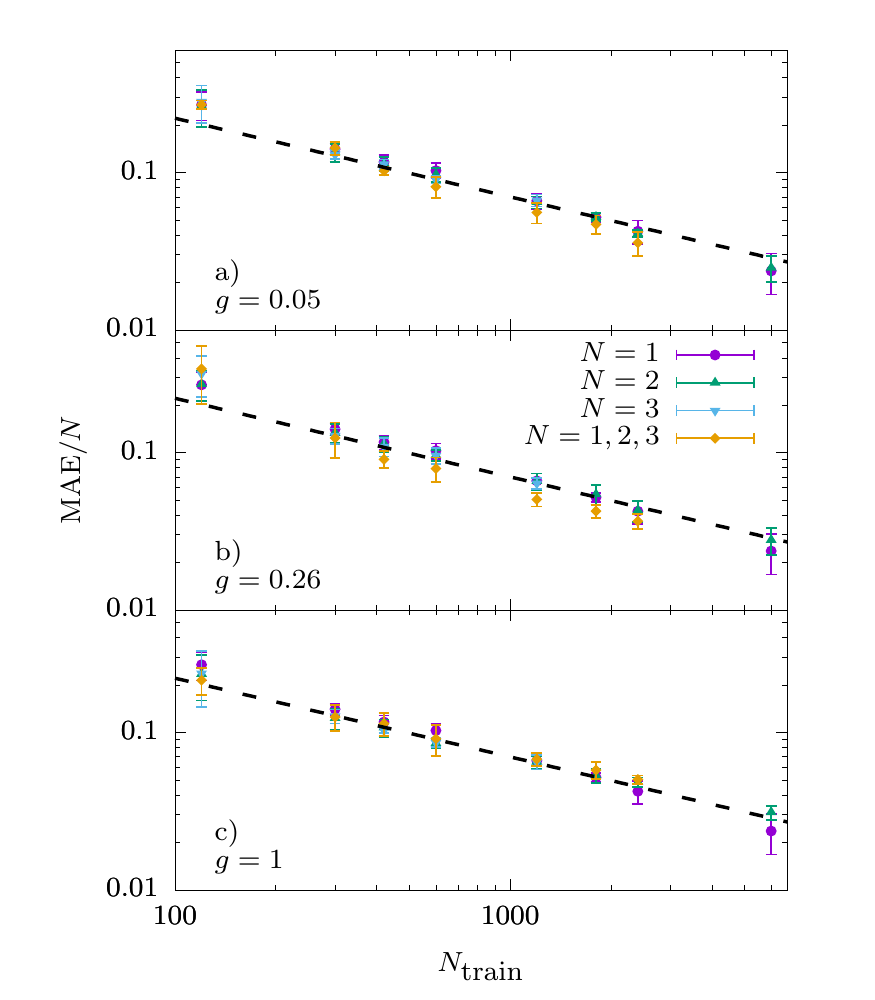}
\caption{Mean absolute error per particle $\mathrm{MAE}/N$, computed  on the test 
ground-state energies, as a function of the number of instances in the training 
set $N_\textrm{train}$. The different symbols correspond to training on homogeneous 
datasets including a unique particle number (either $N=1$, $N=2$, or $N=3$) and 
to combined training and testing on heterogeneous datasets including all three 
particle numbers ($N=1,2,3$). The three panels correspond to different 
interaction strengths $g$. 
The errorbar is the estimated standard deviation of the mean obtained with up to eight independent 
models trained with different pseudo-random numbers. 
The dashed line corresponds to a power-law scaling with $b=0.5$, see text for details. 
\label{fig2}}
\end{figure}

\subsection{Homogeneous datasets}

The neural networks described in the previous section 
are trained to predict the ground-state 
energy of the Hamiltonian~\eqref{Eqhamiltonian}.
We first analyse homogeneous datasets including instances with 
a unique particle number. In this first analysis, the networks are 
trained and tested on the same system size, considering the cases 
$N=1$, $N=2$, and $N=3$ separately.
Since in this analysis the particle number is fixed, we adopt the 
network architecture shown in panel (b) of Fig.~\ref{fig1}, i.e., the 
one that only accepts external potential values as system descriptors.
Three values of the interaction parameter are (separately) considered, 
namely, $g=0.05$, $g=0.26$, and $g=1$. The first choice corresponds to 
the weakly-interacting regime, where the ground-state energies are not far 
from their non-interacting values. The second choice represents an intermediate 
interaction strength, and the third choice is close to the Tonks-Girardeau 
limit where the bosonic ground-state energy approaches the result corresponding 
to (non-interacting) identical fermions.
The learning speed is analysed in Fig.~\ref{fig2}.  The prediction accuracy, as 
measured by the MAE per particle computed on the test set, is plotted as 
a function of the number of instances included in the training 
set $N_\mathrm{train}$. 
It is worth reminding that the test is performed on instances not included in 
the training and in the validation sets. 
In general, one expects a power-law scaling of the prediction accuracy, 
corresponding to 
$\mathrm{MAE}/N\propto N_{\mathrm{train}}^{-b}$~\cite{muller1996numerical,doi:10.1063/1.4964627}, where $b>0$.
Interestingly, we find that the data for all particle numbers and for all 
interaction strengths we consider are consistent with a power-law scaling with 
the same exponent $b=0.5$ (see dashed line in Fig.~\ref{fig2}). These 
results suggest an approximate universal behavior, at least for the 
one-dimensional many-body localized model we address.
While our focus is on the scaling exponent $b$, one notices that the datasets corresponding to different $N$ and $g$ essentially overlap, within the statistical uncertainties. This suggests that also 
the prefactor is, at least approximately, universal.

\subsection{Heterogeneous datasets}

One of our main goals is to implement models that can address different system 
sizes simultaneously. This is achieved via the modified neural-network shown in 
panel (a) of Fig.~\ref{fig1}. This model is fed with an additional descriptor 
representing the particle number $N$, beyond the 1024 speckle potential intensities.
We train and test this model using heterogeneous datasets which include system 
instances with different particle numbers, with equal populations for the 
three $N$ values. As before, training and testing are performed for the same 
interaction parameter, addressing  separately the three values 
we consider.
We stress, however, that in this case the same neural network predicts ground-state 
energies for different particle numbers, while in the previous analysis different 
models were employed for each case. Notably, the MAE per particle follows the 
same power-law scaling with exponent $b=0.5$, as previously found in the analysis 
with separate particle numbers. 
This further supports the statement about an approximately universal behaviour.

%
%
\section{Extrapolation and accelerated learning}
\label{sec5}

\begin{table}[t!]
\centering
\begin{tabular}{|c|c|c|c|c|c|c|}
\hline

\hline
 &\multicolumn{2}{|c|}{$g=$ 0.05} & \multicolumn{2}{|c|}{$g=$ 0.26}  & \multicolumn{2}{|c||}{$g=1$}  \\
\hline
& $R^2$ & MAE & $R^2$  & MAE & $R^2$ & MAE  \\
\hline
\rowcolor[gray]{0.96}
\multicolumn{7}{|c|}{Trained with $N=1,2$} \\
\hline
$N=1$ (1800) & 0.992  & 0.027 & 0.991&0.027 &0.987 & 0.035 \\
\hline
$N=2$ (1800)& 0.994  & 0.047 &0.995 & 0.042& 0.988& 0.065 \\
\hline
$N=3$ (Extrap.) &0.912  &0.299 &0.880&0.366 &0.848 &0.374  \\
\hline
\rowcolor[gray]{0.95}
\multicolumn{7}{|c|}{Accelerated learning for $N=3$} \\
\hline
$N=1$ (1800) & 0.993  & 0.024 & 0.991&0.029 &0.987 & 0.037 \\
\hline
$N=2$ (1800)& 0.992  & 0.046 &0.994 & 0.045& 0.991& 0.057 \\
\hline
$N=3$ (200)  & 0.992 & 0.076 & 0.993 & 0.080 & 0.984 & 0.120  \\
\hline
\hline
\rowcolor[gray]{0.95}
\multicolumn{7}{|c|}{Trained with $N=1,2,3$} \\
\hline
$N=1$ (1200) &0.993  &0.026 & 0.987 &0.039 & 0.980&0.042  \\
\hline
$N=2$ (1200)  & 0.991  & 0.050&0.991 &0.058 &0.992 &0.054  \\
\hline
$N=3$ (1200) &0.995  &0.065 &0.995 &0.070 &0.995 & 0.065 \\
\hline
$N=4$ (Extrap.) & 0.977 &0.172 & 0.920 & 0.313&0.830 &0.481  \\
\hline
\rowcolor[gray]{0.95}
\multicolumn{7}{|c|}{Accelerated learning for $N=4$} \\
\hline
$N=1$ (1200) &0.987  &0.037 & 0.981 &0.046 & 0.980&0.049  \\
\hline
$N=2$ (1200)  & 0.987  & 0.068&0.988 &0.068 &0.987 &0.069  \\
\hline
$N=3$ (1200) &0.991  &0.093 &0.990 &0.099 &0.990 & 0.099 \\
\hline
$N=4$ (200)&0.983 &0.160 & 0.984&0.148 & 0.988 & 0.124  \\
\hline
\end{tabular}
\caption{Performance of the neural network in the test-case considered 
in Figs.~\ref{fig3} and~\ref{fig4}. The coefficient of determination $R^2$ 
and the mean absolute error MAE are reported for three interaction strengths 
$g$, considering networks trained on $N=1,2$ and on $N=1,2,3$. The test 
results are shown for the particle numbers $N$ included in the training 
set (number of training instances in parenthesis), for the extrapolations 
to $N=3$ and to $N=4$, and for the accelerated learning with additional 
large-size instances in the training 
sets.
\label{table1} }
\end{table}

\begin{figure}[t!]
\centering
\subfloat{%
\includegraphics[width=0.65\columnwidth,valign=t]{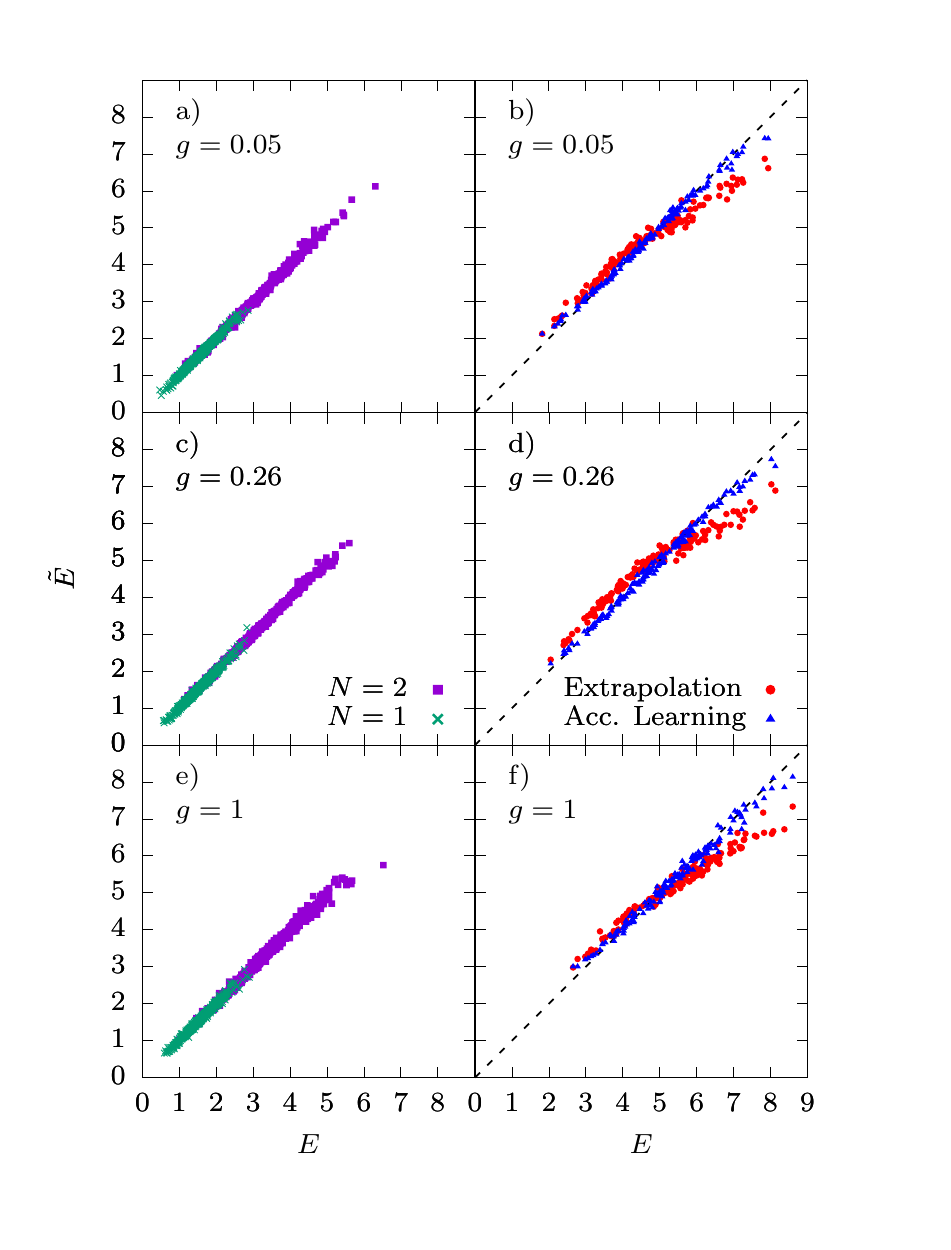}
}
 \hfill
\subfloat{%
\includegraphics[width=0.34\columnwidth,valign=t]{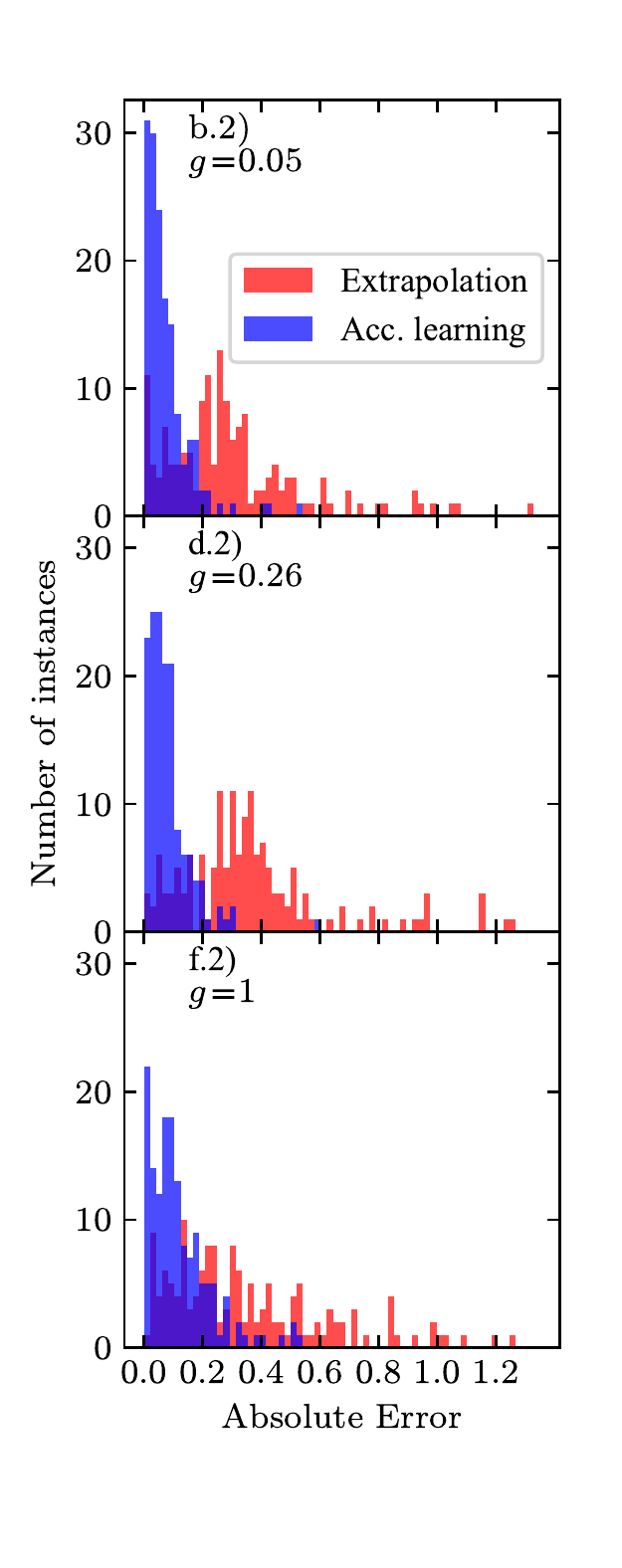}}
\caption{(Left) Ground-state energies $\tilde{E}$, predicted by a neural network 
trained on a heterogeneous dataset, as a function of the 
exact-diagonalization $E$. The training sets include 1800 instances 
for $N=1$ and as many for $N=2$. Panels (a), (c), and (e) report results 
for the systems sizes included in the training set. Panels (b), (d), and (f) 
report the extrapolations to the $N=3$ case, and the accelerated learning 
with 200 additional instances for $N=3$. (Right) Panels (b.2), (d.2), and (f.2) show the distributions of the absolute error, $|\tilde{E}-E|$, for the extrapolations and the accelerated-learning procedure corresponding to panels (b), (d), and (f), respectively. The three rows correspond to 
different interaction strengths $g$.}
\label{fig3}
\end{figure}

The computational cost required to solve many-body problems increases 
exponentially fast with the number of particles. For example, with our 
exact-diagonalization technique the cost increases by a factor 
$\approx 27$ going from the $N=2$ case to the $N=3$ case, as well as 
when going from the $N=3$ to the $N=4$ case. Hence, one expects that 
the datasets one encounters in practical scenarios contain many small 
$N$ instances, and only very few instances for relatively large $N$.
It is therefore natural to wonder (i) if a variable-$N$ neural network can 
perform extrapolations, providing predictions for system sizes larger than 
those included in the training sets, and (ii) if the many small-$N$ instances 
can be used to accelerate the training process for larger $N$, enabling  
the network to provide accurate predictions even when only very few 
training instances are available for the larger system size.
In the following, we address these relevant issues using the variable-$N$ 
architecture shown in panel (a) of Fig.~\ref{fig1}.
First, in Section~\ref{ssn3} we focus on the extrapolation and on the 
accelerated learning of the $N=3$ case, using data for $N=1$ and $N=2$; 
then, in Section~\ref{ssn4} we address the $N=4$ case, where we use data 
for $N=1, 2$, and $3$. Finally, in Section~\ref{sslarge} we consider 
a real-case scenario with much larger databases for lower particle numbers. 

\subsection{Extrapolation and accelerated learning for three particles}
\label{ssn3}

In the first case, a network is trained on a dataset  
including 1800 instances for $N=1$ and as many for $N=2$. 
This network is then used to predict the ground-state energies 
of $N=3$ instances. To quantify the prediction accuracy we consider 
the MAE and the coefficient of determination $R^2$. The corresponding values 
are reported in Table~\ref{table1}. 
For the system sizes included in the training set, namely, $N=1$ and $N=2$, 
the predictions are extremely accurate, corresponding to $R^2 \gtrsim 0.99$.
The high degree of accuracy can be appreciated also in the scatter plots 
of Fig.~\ref{fig3} (panels (a), (c), and (e)), where the predicted energies for the test 
set are plotted as a function of the exact-diagonalization results. Interestingly, 
also the extrapolation to $N=3$ are reasonably accurate, providing 
coefficients of determination $R^2 \gtrsim 0.85$ for all interaction strengths.
The predictions appear to deviate from the exact values mostly in the large 
energy regime (see panels (b), (d), and (f) of Fig.~\ref{fig3}). 
The distributions of absolute errors are shown in panels (b2), (d2), and (f2).
The MAE per particle is around $\mathrm{MAE}/N \simeq 0.1$. 
While this accuracy is remarkable, given that no $N=3$ instance is exploited 
in the training process, it might not be sufficient for practical applications 
of supervised machine learning.
Hence, we analyse the effect of adding to the previous training set just
200 instances for the particle number  $N=3$. We emphasize that the training is performed from scratch using the merged dataset. Interestingly, the combined 
training with the $N=1$, $N=2$, and $N=3$ instances leads to high accuracy for 
all three system sizes. The coefficient of determination is $R^2 \gtrsim 0.99$. 
The MAE per particle for $N=3$ is $\mathrm{MAE}/N \simeq 0.03$, i.e., close to the 
accuracy obtained for $N=1$ and for $N=2$. Notably, the performances on 
the two smaller system sizes do not degrade. For the sake of comparison, 
it is worth noticing that, when the network is trained using only $N=3$ 
instances (see Section~\ref{sec4}), the MAE per particle reached with 
just 200 training instances is approximately an order of magnitude larger.
These results indicate that the combined training with smaller sizes 
provides a boost to the learning process for the larger size, allowing 
the network to reach high accuracy with fewer training instances.

\subsection{Extrapolation and accelerated learning for four particles}
\label{ssn4}
The procedure described above is now extended to $N=4$ systems.
First, a network trained on a dataset including 1200 instances for 
$N=1$, as many for $N=2$ as well as for $N=3$ (corresponding to a 
total of 3600 instances), is used to predict the ground-state energies 
of $N=4$ instances.
\begin{figure}[t!]
\centering
\resizebox{\columnwidth}{!}{
\subfloat{%
\includegraphics[width=0.65\columnwidth,valign=t]{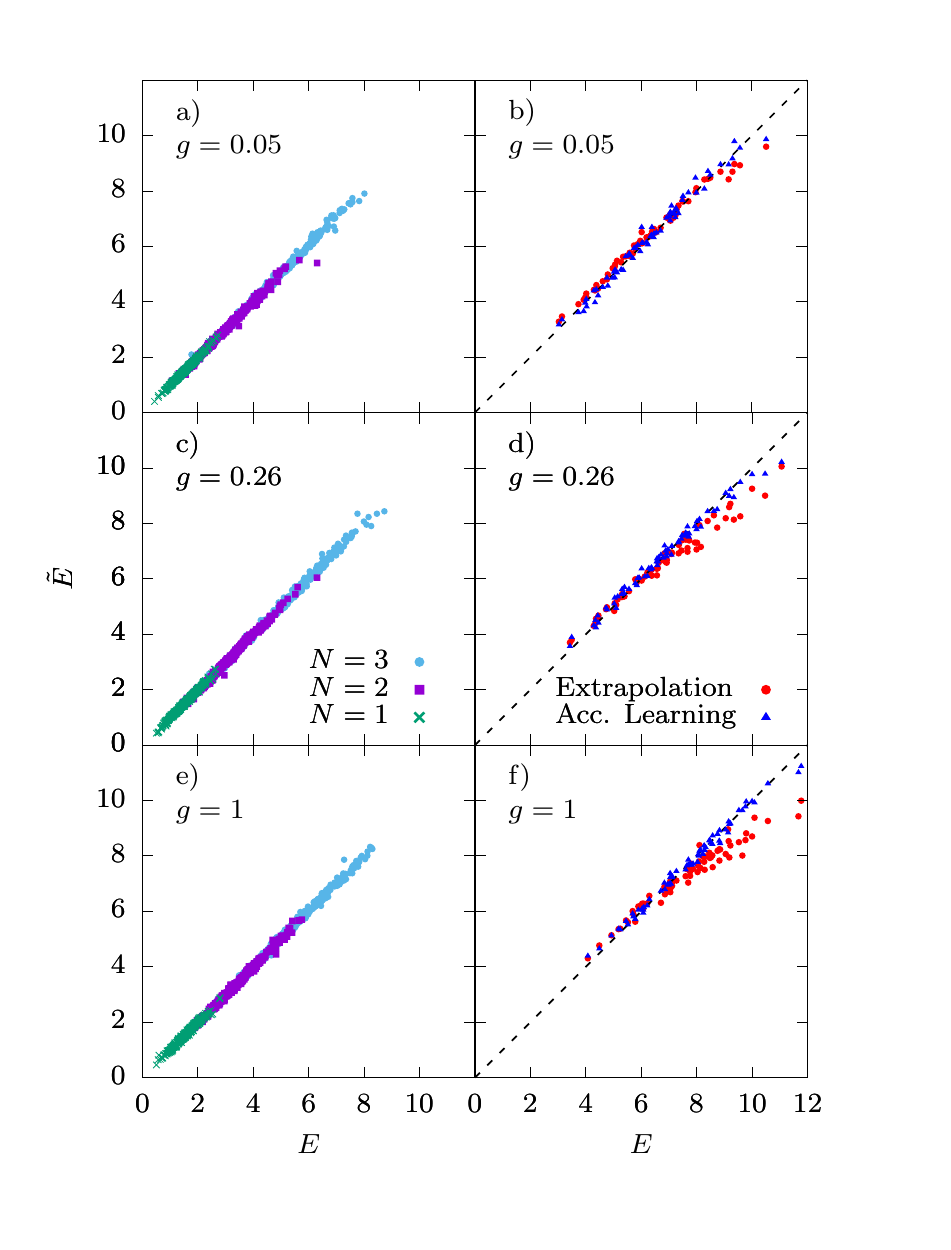}
}\hfill
\subfloat{%
\includegraphics[width=0.34\columnwidth,valign=t]{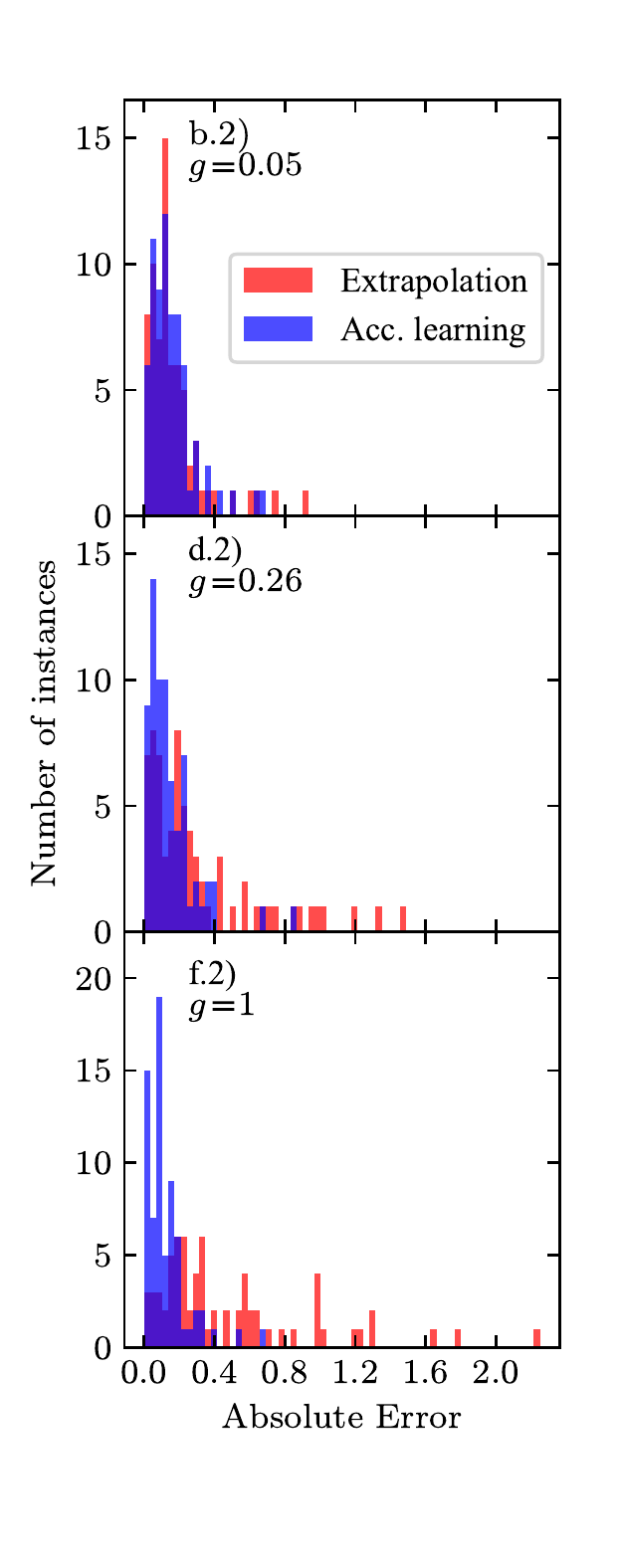}}
}
\caption{(Left) Ground-state energies $\tilde{E}$, predicted by neural networks trained 
on heterogeneous datasets, as a function of the exact-diagonalization results $E$. 
The training sets include 1200 instances for the particle numbers $N=1,2$, 
and $3$, for a total of 3600 instances. Panels (a), (c), and (e) report results for the 
systems sizes included in the training set. Panels (b), (d), and (f) report the 
extrapolations to the $N=4$ case, and the accelerated learning with 200
additional training instances for $N=4$. (Right) Panels (b.2), (d.2), and (f.2) show the distributions of the absolute error, $|\tilde{E}-E|$, for the extrapolations and the accelerated-learning procedure corresponding to panels (b), (d), and (f), respectively. The three rows correspond to different 
interaction strengths $g$.
\label{fig4}}
\end{figure}
The scatter plots of these extrapolations are shown 
in Fig.~\ref{fig4}. In the weak and intermediate interaction regime, the prediction accuracy is remarkably good, 
if one considers that no $N=4$ instance is used in the training process. Specifically, the 
coefficient of determinations are: $R^2\simeq 0.97$ for the weakly interacting 
case $g=0.05$, $R^2\simeq 0.92$ for $g=0.26$, and $R^2\simeq 0.84$ for the 
strongly interacting case $g=1$. 
This suggests that the network is learning how the ground-state energy 
scales with the particle number, at least for the weak and the 
intermediate interactions.
However, the prediction accuracy is not always satisfactory.
Next, we test the efficiency of accelerated learning. We include in the 
previous training set 200 instances for $N=4$. As in Section~\ref{ssn3}, we find consistently accurate results, corresponding to a coefficient of determination 
$R^2 \gtrsim 0.98$ and a $\mathrm{MAE}/N \sim 0.035$ for all interaction 
strengths. For comparison, a network trained only on 200 $N=4$ 
instances (using the model of panel (b) of Fig.~\ref{fig1}) would reach 
$\mathrm{MAE}/N \sim 0.17$ ($R^2\simeq 0.7$). 
Again, this indicates that transfer learning from smaller to larger system 
sizes is effective, allowing one to accelerate the training process for the larger systems.

\subsection{Accelerated learning in a real-case scenario}
\label{sslarge}
\begin{table}[t]
\centering
\begin{tabular}{||c|c|c|c|c|c|c||}
\hline

\hline
 &\multicolumn{2}{|c|}{$g=$ 0.05} & \multicolumn{2}{|c|}{$g=$ 0.26}  & \multicolumn{2}{|c||}{$g=1$}  \\
\hline
& $R^2$ & MAE & $R^2$  & MAE & $R^2$ & MAE  \\
\hline
\rowcolor[gray]{0.96}
\multicolumn{7}{|c|}{Trained with $N=1,2$} \\
\hline
$N=1$ (360000) & 0.9994  & 0.0078 & 0.9994&0.0076 &0.9992& 0.0082 \\
\hline
$N=2$ (30000)& 0.9994  & 0.0138 &0.9986 & 0.0231& 0.9975& 0.0298 \\
\hline
$N=3$ (Extrap.) &0.9641  &0.1889 &0.9335&0.3049 &0.9445 &0.2280  \\
\hline
$N=4$ (Extrap.) &0.8846  &0.4241 &0.7409&0.8045 &0.8427 &0.5164  \\
\hline
\rowcolor[gray]{0.95}
\multicolumn{7}{|c|}{Accelerated learning for $N=3$} \\
\hline
$N=1$ (360000) & 0.9996  & 0.0063 & 0.9993&0.0081 &0.9993 & 0.0081 \\
\hline
$N=2$ (30000)& 0.9996 & 0.0123 &0.9986 & 0.0223& 0.9978& 0.0277\\
\hline
$N=3$ (1500)  & 0.9994 & 0.0219 & 0.9968 & 0.0552 & 0.9928& 0.0798  \\
\hline
\hline
\rowcolor[gray]{0.95}
\multicolumn{7}{|c|}{Trained with $N=1,2,3$ } \\
\hline
$N=1$ (360000) &0.9995  &0.0069 & 0.9993 &0.0081 & 0.9987&0.0111  \\
\hline
$N=2$ (30000)  & 0.9995  & 0.0138&0.9989 &0.0212 &0.9952 &0.0418  \\
\hline
$N=3$ (1200) &0.9993  &0.0234 &0.9975 &0.0534 &0.9904 & 0.0918\\
\hline
$N=4$ (Extrap.) & 0.9934 &0.1140 & 0.9890 & 0.1385&0.9777 &0.1935  \\
\hline
\rowcolor[gray]{0.95}
\multicolumn{7}{|c|}{Accelerated learning for $N=4$} \\
\hline
$N=1$ (360000) & 0.9994 &0.0074 &0.9994 & 0.0075 & 0.9992&0.0087  \\
\hline
$N=2$ (30000)  & 0.9992  &0.0165 & 0.9987 & 0.0226 &0.9974 &0.0301 \\
\hline
$N=3$ (1200) &0.9988 & 0.0346& 0.9974 &0.0523 &0.9938 & 0.0789 \\
\hline
$N=4$ (200)& 0.9987&0.0438 & 0.9925 & 0.1100 & 0.9882 & 0.1459  \\
\hline
\end{tabular}
\caption{Performance of the neural network in the real-case scenario. 
The coefficient of determination $R^2$ and the mean absolute error 
MAE are reported for three interaction strengths $g$, considering 
networks trained on $N=1,2$ and on $N=1,2,3$. The test results are 
shown for the particle numbers $N$ included in the training set 
(number of training instances in parenthesis), for the extrapolations 
to $N=3$ and to $N=4$, and for the accelerated learning. 
\label{table2} }
\end{table}
Since the computational cost of solving many-body instances 
increases exponentially fast with the systems size, in practical 
applications of supervised learning the training sets inevitably 
contain significantly fewer instances for the larger particle numbers. 
Here, we analyse the efficiency of the accelerated learning with 
the typical training dataset one would encounter in a 
real-case scenario. Specifically, this dataset includes 360000 
instances for $N=1$, 30000 for $N=2$, 1200 for $N=3$, and 200 for $N=4$. 
It is worth noticing that more computational time is invested in 
the larger particle numbers, since one expects that larger systems 
provide more information about the scaling of the ground-state 
properties with the system size.
The performance of the extrapolations and of the accelerated learning 
is summarized in Table~\ref{table2}, where we report the MAE and 
the coefficient of determination. 
Interestingly, the extrapolations are significantly more accurate 
and more consistent than those reported in Section~\ref{ssn4}, which were based on fewer 
training instances (see Table~\ref{table1}). In particular, the 
extrapolations to $N=4$, based on training instances for $N=1$, 
$N=2$, and $N=3$, reach $R^2\gtrsim 0.97$. 
Instead, when the training set includes only $N=1$ and $N=2$ instances, the extrapolations to $N=3$ and, even more, those to $N=4$, are less accurate, reaching coefficient of determinations up to $\approx 25\%$ lower.
This suggests that, 
only when the training set includes several system sizes and is sufficiently copious, the network can 
learn to accurately scale the predictions to larger particle numbers.
However, a definitive assessment on the extrapolation accuracy would require testing even larger particle numbers, which are currently out of reach for the system under investigation.
Notably, we observe that including in the training set 200 instances for $N=4$ (beyond the $N=1$, $N=2$, and $N=3$ instances) is sufficient to further 
improve the accuracy. Again, this indicates that the network 
is capable of transferring the knowledge acquired on smaller system sizes, 
using it to drastically accelerate the learning of larger-system properties, leading to systematically accurate predictions.

 \section{Summary and conclusions}
 \label{sec6}

We have addressed the supervised learning of the (few) dirty boson problem, 
considering a specific Hamiltonian which has already been implemented in cold-atom experiments. 
The training and the test sets have been produced via an exact 
diagonalization technique, avoiding the uncontrolled approximations often 
employed in analogous studies on the supervised learning of quantum systems.
This limited our analysis to relatively small systems, specifically, up 
to four bosons.
These datasets are made publicly available at Ref.~\cite{zenodo} to support future comparative studies on the supervised training of deep neural networks.
Our findings indicate that the learning curve, in terms of accuracy of 
ground-state energy predictions versus number of training instances, is 
approximately universal for different particle numbers and for different 
interaction strengths. An interesting open question is whether a completely different neural network architecture can provide even faster learning, therefore breaking the observed universal behavior.

The artificial neural network we introduced can be trained and tested on 
heterogeneous datasets including instances with different particle numbers. 
This is achieved by combining a convolutional architecture  
which can address disordered fields of variable 
spatial extent, with an additional descriptor that explicitly represents the 
particle number. This descriptor is fed to the final dense layers, bypassing 
the convolutional part. This detail constitutes a relevant innovative aspect 
of our architecture. Our analysis demonstrates that this network provides 
accurate ground-state energy predictions, independently of the particle number, 
at least within the system sizes  we considered. Furthermore, it allows one 
to attempt extrapolations to particle numbers larger than those included in the training set. 
The accuracy of these extrapolations is variable, and it improves when more instances and larger particle numbers are included in the training set.
Notably, the learning of relatively large systems can 
be accelerated and made consistently accurate using heterogeneous training sets including many small-size 
instances and only a small amount of large-size instances. This represents 
the typical scenario, given the rapidly growing computational cost of 
solving quantum models. This strategy is somewhat analogous to the transfer 
learning protocols commonly employed in the field of image analysis, whereby 
deep neural networks pre-trained on large datasets -- relevant examples are 
the ResNet~\cite{he2016deep} and the VGG models~\cite{simonyan2014very} -- are 
then specialized on the desired classification task using much smaller samples. 
Here we implemented transfer learning from small to larger particle numbers 
using heterogeneous datasets.

In future work, it would be interesting to further explore the universality 
of the learning curve, considering setups with different models of disorder, 
interatomic potentials, geometries, or particle statistics. Furthermore, it 
would be important to extend our analysis to larger particle numbers, 
possibly in combination with different computational techniques, such as, e.g., 
quantum Monte Carlo simulations. Additionally, it would be interesting to consider other physical quantities, e.g., the different contributions to the ground-state energy. Also in this case, model training might be accelerated using pre-trained models in a transfer-learning strategy.
As a future perspective, one can envision the use of cold-atom experiments
as quantum simulators to produce the datasets required to train neural 
networks for computationally intractable models. In typical experiments, the number of atoms varies due to three-body recombinations, but the deterministic preparation of few-atoms systems with well-controlled atom number has recently been achieved ~\cite{Serwane336,Wenz457}. In any case, it is convenient to use the particle number as an explicit system descriptor. This allows performing supervised learning with heterogeneous datasets obtained from different experimental setups. We argue that flexible 
neural-network architecture and transfer learning strategies shall play a 
critical role in the practical applications of cold-atom quantum simulators.
Furthermore, the application of classical machine-learning techniques to study quantum systems is being developed in parallel with quantum machine-learning algorithms, which may outperform in both classical and quantum tasks~\cite{Lamata_2020,Biamonte2017,PhysRevLett.122.040504,Dunjko_2018,Ciliberto2018}.

\section*{Acknowledgements}
We acknowledge useful discussions with S. Cantori, A. Dauphin, 
F. Isaule and I. Morera.


\paragraph{Funding information}
S. P. acknowledges financial support from 
the FAR2018 project titled ``Supervised machine learning for quantum
matter and computational docking'' of the University of Camerino and 
from the Italian MIUR under the project PRIN2017 CEnTraL 20172H2SC4. 
This work has been partially supported by MINECO (Spain) Grant 
No. FIS2017-87534-P. We acknowledge financial support from Secretaria d'Universitats i Recerca del Departament d'Empresa i Coneixement de la Generalitat de Catalunya, co-funded by the European Union Regional Development Fund within the ERDF Operational Program of Catalunya (project QuantumCat, ref. 001-P-001644). S. P. also acknowledges the CINECA award under the ISCRA initiative, for the availability of high performance computing resources and support.
\bibliography{Ref.bib}

\begin{thebibliography}{10}
\providecommand{\url}[1]{\texttt{#1}}
\providecommand{\urlprefix}{URL }
\expandafter\ifx\csname urlstyle\endcsname\relax
  \providecommand{\doi}[1]{doi:\discretionary{}{}{}#1}\else
  \providecommand{\doi}{doi:\discretionary{}{}{}\begingroup
  \urlstyle{rm}\Url}\fi
\providecommand{\eprint}[2][]{\url{#2}}

\bibitem{carleo2019machine}
G.~Carleo, I.~Cirac, K.~Cranmer, L.~Daudet, M.~Schuld, N.~Tishby,
  L.~Vogt-Maranto and L.~Zdeborov{\'a},
\newblock \emph{Machine learning and the physical sciences},
\newblock Rev. Mod. Phys. \textbf{91}(4), 045002 (2019),
\newblock \doi{10.1103/RevModPhys.91.045002}.

\bibitem{carrasquilla2020machine}
J.~Carrasquilla,
\newblock \emph{Machine learning for quantum matter},
\newblock Advances in Physics: X \textbf{5}(1), 1797528 (2020),
\newblock \doi{10.1080/23746149.2020.1797528}.

\bibitem{blank1995neural}
T.~B. Blank, S.~D. Brown, A.~W. Calhoun and D.~J. Doren,
\newblock \emph{Neural network models of potential energy surfaces},
\newblock J. Chem. Phys. \textbf{103}(10), 4129 (1995),
\newblock \doi{10.1063/1.469597}.

\bibitem{behler2007generalized}
J.~Behler and M.~Parrinello,
\newblock \emph{Generalized neural-network representation of high-dimensional
  potential-energy surfaces},
\newblock Phys. Rev. Lett. \textbf{98}(14), 146401 (2007),
\newblock \doi{10.1103/PhysRevLett.98.146401}.

\bibitem{behler2011neural}
J.~Behler,
\newblock \emph{Neural network potential-energy surfaces in chemistry: a tool
  for large-scale simulations},
\newblock Phys. Chem. Chem. Phys. \textbf{13}(40), 17930 (2011),
\newblock \doi{10.1039/C1CP21668F}.

\bibitem{bartok2017machine}
A.~P. Bart{\'o}k, S.~De, C.~Poelking, N.~Bernstein, J.~R. Kermode,
  G.~Cs{\'a}nyi and M.~Ceriotti,
\newblock \emph{Machine learning unifies the modeling of materials and
  molecules},
\newblock Sci. Adv. \textbf{3}(12), e1701816 (2017),
\newblock \doi{10.1126/sciadv.1701816}.

\bibitem{zhang2018end}
L.~Zhang, J.~Han, H.~Wang, W.~Saidi, R.~Car and E.~Weinan,
\newblock \emph{End-to-end symmetry preserving inter-atomic potential energy
  model for finite and extended systems},
\newblock In S.~Bengio, H.~Wallach, H.~Larochelle, K.~Grauman, N.~Cesa-Bianchi
  and R.~Garnett, eds., \emph{Adv. Neural Inf. Process. Syst.}, pp. 4436--4446.
  Curran Associates, Inc. (2018).

\bibitem{snyder2012finding}
J.~C. Snyder, M.~Rupp, K.~Hansen, K.-R. M{\"u}ller and K.~Burke,
\newblock \emph{Finding density functionals with machine learning},
\newblock Phys. Rev. Lett. \textbf{108}(25), 253002 (2012),
\newblock \doi{10.1103/PhysRevLett.108.253002}.

\bibitem{li2016understanding}
L.~Li, J.~C. Snyder, I.~M. Pelaschier, J.~Huang, U.-N. Niranjan, P.~Duncan,
  M.~Rupp, K.-R. M{\"u}ller and K.~Burke,
\newblock \emph{Understanding machine-learned density functionals},
\newblock Int. J. Quantum Chem. \textbf{116}(11), 819 (2016),
\newblock \doi{10.1002/qua.25040}.

\bibitem{brockherde2017bypassing}
F.~Brockherde, L.~Vogt, L.~Li, M.~E. Tuckerman, K.~Burke and K.-R. M{\"u}ller,
\newblock \emph{{Bypassing the Kohn-Sham equations with machine learning}},
\newblock Nat. Commun. \textbf{8}(1), 872 (2017),
\newblock \doi{10.1038/s41467-017-00839-3}.

\bibitem{ryczko2019deep}
K.~Ryczko, D.~A. Strubbe and I.~Tamblyn,
\newblock \emph{Deep learning and density-functional theory},
\newblock Phys. Rev. A \textbf{100}(2), 022512 (2019),
\newblock \doi{10.1103/PhysRevA.100.022512}.

\bibitem{PhysRevLett.125.076402}
J.~R. Moreno, G.~Carleo and A.~Georges,
\newblock \emph{Deep learning the hohenberg-kohn maps of density functional
  theory},
\newblock Phys. Rev. Lett. \textbf{125}, 076402 (2020),
\newblock \doi{10.1103/PhysRevLett.125.076402}.

\bibitem{custodio2019artificial}
C.~A. Cust{\'o}dio, {\'E}.~R. Filletti and V.~V. Fran{\c{c}}a,
\newblock \emph{Artificial neural networks for density-functional optimizations
  in fermionic systems},
\newblock Sci. Rep. \textbf{9}(1), 1886 (2019),
\newblock \doi{10.1038/s41598-018-37999-1}.

\bibitem{hansen2013assessment}
K.~Hansen, G.~Montavon, F.~Biegler, S.~Fazli, M.~Rupp, M.~Scheffler, O.~A.
  Von~Lilienfeld, A.~Tkatchenko and K.-R. M{\"u}ller,
\newblock \emph{Assessment and validation of machine learning methods for
  predicting molecular atomization energies},
\newblock J. Chem. Theory Comput. \textbf{9}(8), 3404 (2013),
\newblock \doi{10.1021/ct400195d}.

\bibitem{schutt2014represent}
K.~T. Sch{\"u}tt, H.~Glawe, F.~Brockherde, A.~Sanna, K.~R. M{\"u}ller and
  E.~K.~U. Gross,
\newblock \emph{How to represent crystal structures for machine learning:
  Towards fast prediction of electronic properties},
\newblock Phys. Rev. B \textbf{89}(20), 205118 (2014),
\newblock \doi{10.1103/PhysRevB.89.205118}.

\bibitem{hansen2015machine}
K.~Hansen, F.~Biegler, R.~Ramakrishnan, W.~Pronobis, O.~A. Von~Lilienfeld,
  K.-R. M{\"u}ller and A.~Tkatchenko,
\newblock \emph{Machine learning predictions of molecular properties: Accurate
  many-body potentials and nonlocality in chemical space},
\newblock J. Phys. Chem. Lett. \textbf{6}(12), 2326 (2015),
\newblock \doi{10.1021/acs.jpclett.5b00831}.

\bibitem{ballester2010machine}
P.~J. Ballester and J.~B. Mitchell,
\newblock \emph{A machine learning approach to predicting protein--ligand
  binding affinity with applications to molecular docking},
\newblock Bioinformatics \textbf{26}(9), 1169 (2010),
\newblock \doi{10.1093/bioinformatics/btq112}.

\bibitem{khamis2015machine}
M.~A. Khamis, W.~Gomaa and W.~F. Ahmed,
\newblock \emph{Machine learning in computational docking},
\newblock Artif. Intell. Med. \textbf{63}(3), 135 (2015),
\newblock \doi{10.1016/j.artmed.2015.02.002}.

\bibitem{jimenez2018k}
J.~Jim{\'e}nez, M.~Skalic, G.~Martinez-Rosell and G.~De~Fabritiis,
\newblock \emph{K deep: Protein--ligand absolute binding affinity prediction
  via 3d-convolutional neural networks},
\newblock J. Chem. Inf. Model. \textbf{58}(2), 287 (2018),
\newblock \doi{10.1021/acs.jcim.7b00650}.

\bibitem{funahashi1989approximate}
K.-I. Funahashi,
\newblock \emph{On the approximate realization of continuous mappings by neural
  networks},
\newblock Neural Netw. \textbf{2}(3), 183 (1989),
\newblock \doi{10.1016/0893-6080(89)90003-8}.

\bibitem{caruana1997multitask}
R.~Caruana,
\newblock \emph{Multitask learning},
\newblock Mach. Learn. \textbf{28}(1), 41 (1997),
\newblock \doi{10.1023/A:1007379606734}.

\bibitem{behler2016perspective}
J.~Behler,
\newblock \emph{{Perspective: Machine learning potentials for atomistic
  simulations}},
\newblock J. Chem. Phys. \textbf{145}(17), 170901 (2016),
\newblock \doi{10.1063/1.4966192}.

\bibitem{mills2019extensive}
K.~Mills, K.~Ryczko, I.~Luchak, A.~Domurad, C.~Beeler and I.~Tamblyn,
\newblock \emph{Extensive deep neural networks for transferring small scale
  learning to large scale systems},
\newblock Chem. Sci. \textbf{10}(15), 4129 (2019),
\newblock \doi{10.1039/C8SC04578J}.

\bibitem{jungsize}
H.~Jung, S.~Stocker, C.~Kunkel, H.~Oberhofer, B.~Han, K.~Reuter and J.~T.
  Margraf,
\newblock \emph{Size-extensive molecular machine learning with global
  representations},
\newblock Chem Systems Chem \textbf{2}, e1900052 (2020),
\newblock \doi{10.1002/syst.201900052}.

\bibitem{saraceni2020scalable}
N.~Saraceni, S.~Cantori and S.~Pilati,
\newblock \emph{Scalable neural networks for the efficient learning of
  disordered quantum systems},
\newblock Phys. Rev. E \textbf{102}, 033301 (2020),
\newblock \doi{10.1103/PhysRevE.102.033301}.

\bibitem{faber2017prediction}
F.~A. Faber, L.~Hutchison, B.~Huang, J.~Gilmer, S.~S. Schoenholz, G.~E. Dahl,
  O.~Vinyals, S.~Kearnes, P.~F. Riley and O.~A. Von~Lilienfeld,
\newblock \emph{Prediction errors of molecular machine learning models lower
  than hybrid dft error},
\newblock J. of Chem. Theory Comput. \textbf{13}(11), 5255 (2017),
\newblock \doi{10.1021/acs.jctc.7b00577}.

\bibitem{aspect2009anderson}
A.~Aspect and M.~Inguscio,
\newblock \emph{Anderson localization of ultracold atoms},
\newblock Phys. Today \textbf{62}(8), 30 (2009),
\newblock \doi{10.1063/1.3206092}.

\bibitem{fisher1989boson}
M.~P. Fisher, P.~B. Weichman, G.~Grinstein and D.~S. Fisher,
\newblock \emph{Boson localization and the superfluid-insulator transition},
\newblock Phys. Rev. B \textbf{40}(1), 546 (1989),
\newblock \doi{10.1103/PhysRevB.40.546}.

\bibitem{PhysRevB.80.104515}
G.~M. Falco, T.~Nattermann and V.~L. Pokrovsky,
\newblock \emph{Weakly interacting bose gas in a random environment},
\newblock Phys. Rev. B \textbf{80}, 104515 (2009),
\newblock \doi{10.1103/PhysRevB.80.104515}.

\bibitem{PhysRevLett.98.170403}
P.~Lugan, D.~Cl\'ement, P.~Bouyer, A.~Aspect, M.~Lewenstein and
  L.~Sanchez-Palencia,
\newblock \emph{Ultracold bose gases in 1d disorder: From lifshits glass to
  bose-einstein condensate},
\newblock Phys. Rev. Lett. \textbf{98}, 170403 (2007),
\newblock \doi{10.1103/PhysRevLett.98.170403}.

\bibitem{PhysRevA.100.013603}
P.~Mujal, A.~Polls, S.~Pilati and B.~Juli\'a-D\'{\i}az,
\newblock \emph{Few-boson localization in a continuum with speckle disorder},
\newblock Phys. Rev. A \textbf{100}, 013603 (2019),
\newblock \doi{10.1103/PhysRevA.100.013603}.

\bibitem{roati2008anderson}
G.~Roati, C.~D'Errico, L.~Fallani, M.~Fattori, C.~Fort, M.~Zaccanti,
  G.~Modugno, M.~Modugno and M.~Inguscio,
\newblock \emph{{Anderson localization of a non-interacting Bose--Einstein
  condensate}},
\newblock Nature \textbf{453}(7197), 895 (2008),
\newblock \doi{10.1038/nature07071}.

\bibitem{billy2008direct}
J.~Billy, V.~Josse, Z.~Zuo, A.~Bernard, B.~Hambrecht, P.~Lugan, D.~Cl{\'e}ment,
  L.~Sanchez-Palencia, P.~Bouyer and A.~Aspect,
\newblock \emph{{Direct observation of Anderson localization of matter waves in
  a controlled disorder}},
\newblock Nature \textbf{453}(7197), 891 (2008),
\newblock \doi{10.1038/nature07000}.

\bibitem{ley1989specklhunte}
J.~Huntley,
\newblock \emph{Speckle photography fringe analysis: assessment of current
  algorithms},
\newblock Appl. Opt. \textbf{28}(20), 4316 (1989),
\newblock \doi{10.1364/AO.28.004316}.

\bibitem{PhysRevA.73.013606}
M.~Modugno,
\newblock \emph{Collective dynamics and expansion of a bose-einstein condensate
  in a random potential},
\newblock Phys. Rev. A \textbf{73}, 013606 (2006),
\newblock \doi{10.1103/PhysRevA.73.013606}.

\bibitem{plodzien2018numerically}
M.~P{\l}odzie{\'n}, D.~Wiater, A.~Chrostowski and T.~Sowi{\'n}ski,
\newblock \emph{Numerically exact approach to few-body problems far from a
  perturbative regime}  (2018),
\newblock \href{https://arxiv.org/abs/1803.08387}{arXiv:1803.08387}.

\bibitem{pmtphdthesis}
P.~Mujal,
\newblock
  \emph{\href{http://www.ecm.ub.es/~bruno/works/PMT_phD_Thesis_book.pdf}{Interacting
  ultracold few-boson systems}},
\newblock Ph.D. thesis, Universitat de Barcelona (2019).

\bibitem{zenodo}
P.~Mujal, A.~Mart\'{i}nez~Miguel, A.~Polls, B.~Juliá-Díaz and S.~Pilati,
\newblock \emph{Database used in the analysis},
\newblock Zenodo  (2020),
\newblock \doi{10.5281/zenodo.4058492}.

\bibitem{nielsen2015neural}
M.~A. Nielsen,
\newblock \emph{\href{http://neuralnetworksanddeeplearning.com}{Neural Networks
  and Deep Learning}},
\newblock Determination Press (2015).

\bibitem{kingma2014adam}
D.~P. Kingma and J.~Ba,
\newblock \emph{Adam: A method for stochastic optimization}  (2014),
\newblock \href{https://arXiv.org/abs/1412.6980}{arXiv:1412.6980}.

\bibitem{chollet2015keras}
F.~Chollet \emph{et~al.},
\newblock \emph{Keras},
\newblock \url{https://keras.io} (2015).

\bibitem{muller1996numerical}
K.-R. M{\"u}ller, M.~Finke, N.~Murata, K.~Schulten and S.-i. Amari,
\newblock \emph{A numerical study on learning curves in stochastic multilayer
  feedforward networks},
\newblock Neural Computation \textbf{8}(5), 1085 (1996),
\newblock \doi{10.1162/neco.1996.8.5.1085}.

\bibitem{doi:10.1063/1.4964627}
B.~Huang and O.~A. von Lilienfeld,
\newblock \emph{Communication: Understanding molecular representations in
  machine learning: The role of uniqueness and target similarity},
\newblock J. Chem. Phys. \textbf{145}(16), 161102 (2016),
\newblock \doi{10.1063/1.4964627}.

\bibitem{he2016deep}
K.~He, X.~Zhang, S.~Ren and J.~Sun,
\newblock \emph{Deep residual learning for image recognition},
\newblock In \emph{Proc. IEEE Int. Conf. Comput. Vis.}, pp. 770--778,
\newblock \doi{10.1109/CVPR.2016.90} (2016).

\bibitem{simonyan2014very}
K.~Simonyan and A.~Zisserman,
\newblock \emph{Very deep convolutional networks for large-scale image
  recognition}  (2014),
\newblock \href{https://arxiv.org/abs/1409.1556}{arXiv:1409.1556}.

\bibitem{Serwane336}
F.~Serwane, G.~Z{\"u}rn, T.~Lompe, T.~B. Ottenstein, A.~N. Wenz and S.~Jochim,
\newblock \emph{Deterministic preparation of a tunable few-fermion system},
\newblock Science \textbf{332}(6027), 336 (2011),
\newblock \doi{10.1126/science.1201351}.

\bibitem{Wenz457}
A.~N. Wenz, G.~Z{\"u}rn, S.~Murmann, I.~Brouzos, T.~Lompe and S.~Jochim,
\newblock \emph{From few to many: Observing the formation of a fermi sea one
  atom at a time},
\newblock Science \textbf{342}(6157), 457 (2013),
\newblock \doi{10.1126/science.1240516}.

\bibitem{Lamata_2020}
L.~Lamata,
\newblock \emph{Quantum machine learning and quantum biomimetics: A
  perspective},
\newblock Machine Learning: Science and Technology \textbf{1}(3), 033002
  (2020),
\newblock \doi{10.1088/2632-2153/ab9803}.

\bibitem{Biamonte2017}
J.~Biamonte, P.~Wittek, N.~Pancotti, P.~Rebentrost, N.~Wiebe and S.~Lloyd,
\newblock \emph{Quantum machine learning},
\newblock Nature \textbf{549}(7671), 195 (2017),
\newblock \doi{10.1038/nature23474}.

\bibitem{PhysRevLett.122.040504}
M.~Schuld and N.~Killoran,
\newblock \emph{Quantum machine learning in feature hilbert spaces},
\newblock Phys. Rev. Lett. \textbf{122}, 040504 (2019),
\newblock \doi{10.1103/PhysRevLett.122.040504}.

\bibitem{Dunjko_2018}
V.~Dunjko and H.~J. Briegel,
\newblock \emph{Machine learning {\&} artificial intelligence in the quantum
  domain: a review of recent progress},
\newblock Reports on Progress in Physics \textbf{81}(7), 074001 (2018),
\newblock \doi{10.1088/1361-6633/aab406}.

\bibitem{Ciliberto2018}
C.~Ciliberto, M.~Herbster, A.~D. Ialongo, M.~Pontil, A.~Rocchetto, S.~Severini
  and L.~Wossnig,
\newblock \emph{Quantum machine learning: a classical perspective},
\newblock Proc. R. Soc. A \textbf{474}, 20170551 (2018),
\newblock \doi{10.1098/rspa.2017.0551}.

\end{thebibliography}

\nolinenumbers

\end{document}